\documentclass[prb,twocolumn]{revtex4}
\usepackage{graphicx}
\usepackage{bm}
\begin{document}

\title{A spin-polarized scheme for obtaining quasi-particle energies 
within the density functional theory}

\author{ B. Barbiellini and A. Bansil}

\affiliation{Physics Department, Northeastern University, Boston MA 02115}


\begin{abstract}

We discuss an efficient scheme for obtaining spin-polarized quasi-particle 
excitation energies within the general framework of the density functional 
theory (DFT). Our approach is to correct the DFT eigenvalues via the 
electrostatic energy of a majority or minority spin electron resulting 
from its interaction with the associated exchange and correlation holes by 
using appropriate spin-resolved pair correlation functions. A version of 
the method for treating systems with localized orbitals, including the 
case of partially filled metallic bands, is considered. Illustrative 
results on Cu are presented.

\end{abstract}

\maketitle

\section{Introduction}

As spectroscopic data using a variety of experimental techniques is 
becoming available at an ever increasing pace, there is great need to 
develop efficient methods for calculating electronic excitation 
energies
in wide classes of materials. The GW approximation (GWA) has 
been the 
traditional route for addressing this problem \cite{review_GW}. 
The GWA however is computationally quite intensive, 
which makes the routine application of 
the GWA in electronic structure computations difficult and for this 
reason GWA results have to date been limited largely to relatively simple 
systems. 

With this motivation, in a previous study~\cite{bba04}, 
we have attempted to develop a 
simplified scheme for obtaining quasi-particle energies by correcting the 
eigenvalues $\epsilon_n$ given by the density functional theory 
(DFT)\cite{review_dft} with a 
state-dependent correction $\Delta_n$ associated with the self-energy 
$\Sigma_{xc}$ of the exchange-correlation hole surrounding each electron. 
The results are similar to those obtained within other extensions of the 
DFT \cite{fritsche86,fritsche93,freeman94,remed99,reed00,sankey00,sankey01}, 
which have been shown to describe the energy gaps in semiconductors with 
an accuracy comparable to that of the GWA. 
The purpose of the present work is to generalize the unpolarized case 
considered in Ref.~\onlinecite{bba04} to include spin-polarization 
effects. We also address the treatment of self-interaction corrections in 
systems with localized $d$ and $f$ orbitals and suggest a scheme for 
handling metallic systems with partially filled bands of $d$ or $f$ 
character. 

This article is organized as follows. The 
introduction is followed in Section~II by a brief explanation of 
the relevant pair-correlation functions and their role in 
evaluating self-energies. The spin-resolution of the exchange and 
correlation energies is developed in Section~III. A somewhat 
more general formulation of our scheme is taken up in Section~IV which 
also presents illustrative results in Cu. Finally, Section~V makes a few 
concluding remarks.

\section{Role of pair-correlation functions}

We consider the elements
\begin{equation}
\pi({\bf r}_1,{\bf r}_2)
=N(N-1)\int |\Psi({\bf r}_1,...,{\bf r}_N)|^2~
d{\bf r}_{3}...d{\bf r}_{N},
\label{eq_pi}
\end{equation}
of the two-particle density matrix, where $N$ is the total number of
electrons and $\Psi$ is the ground state wavefunction of the many body
electronic system. The pair correlation function
$g({\bf r}_1,{\bf r}_2)$ is then defined via the relation
\begin{eqnarray}
\nonumber
\pi({\bf r}_1,{\bf r}_2)&=&
n({\bf r}_1)n({\bf r}_2)g({\bf r}_1,{\bf r}_2)\\
&=&n({\bf r}_1)n({\bf r}_2)(1+C({\bf r}_1,{\bf r}_2)),
\label{eq_g}
\end{eqnarray}
where $n({\bf r})$ denotes the electron density and the correlation
factor $C({\bf r}_1,{\bf r}_2)$ is defined by the second equality. For an
uncorrelated system $g({\bf r}_1,{\bf r}_2) \rightarrow 1$, or
equivalently, $C({\bf r}_1,{\bf r}_2) \rightarrow 0$.

Two effects must now be considered, namely, exchange and correlation. The
antisymmetry of the many electron wavefunction $\Psi$ prevents electrons of
the same spin to come too close to one another and creates an 'exchange
hole' around each electron. The associated exchange energy is:
\begin{equation}
E_{x}=\int d{\bf r}~n({\bf r})\epsilon_{x}({\bf r})~,
\end{equation}
where
\begin{equation}
\epsilon_{x}({\bf r})=
\frac{1}{2}
\int d{\bf r'} \frac{
n({\bf r'})
C_{x}({\bf r},{\bf r'})}
{|{\bf r}-{\bf r'}|}
\label{ex}
\end{equation}
is the exchange energy per particle and $C_x$ is the exchange contribution
to the pair correlation $C({\bf r},{\bf r'})$ of Eq.~\ref{eq_g}.
The factor of ${1/2}$ accounts for double counting of the electron-electron
interaction.
Thus,
the exchange energy per particle $\epsilon_{x}({\bf r})$ can readily be
interpreted as the Coulomb interaction of an electron at ${\bf r}$ with
its exchange hole. For the homogeneous electron gas (HEG) of density $n$,
$C_x$ and $\epsilon_x$ have well-known analytical expressions
$C_x^{HEG}(n)$ and $\epsilon_x^{HEG}(n)$ and, in particular, the
familiar Hartree-Fock-Slater (HFS) potential is given by \cite{hfs}
\begin{equation}
v_{S}({\bf r})=2\epsilon_x^{HEG}(n({\bf r})).
\label{eq_hfs}
\end{equation}
Here the factor of $2$ arises from the fact that variations with respect
to a trial wavefunction of the form of a Slater determinant yield a
potential term in one-particle equations which is twice as large as the
corresponding energy term $\epsilon_x$. The DFT within the local density 
approximation (LDA) provides a remarkably
similar scheme in which the electron density $n({\bf r})$ is treated as
the variational parameter\cite{ks}. The exchange hole is still described
by the pair correlation function between like spin electrons, but the
exchange potential
\begin{equation}
v_x({\bf r})=\frac{\delta E_x}{\delta n({\bf r})}
\end{equation}
differs from the HFS potential by a factor of $2/3$.

Coulomb repulsion between electrons is an additional effect responsible
for electron motions to become correlated. As a result, there also appears
a 'correlation hole' around each electron, which is dominated by electrons
of opposite spin, since the like spin electrons are already excluded via
the exchange hole. The correlation hole is generally less deep compared to
the exchange hole, especially at high electron densities. Note that the
exchange hole goes to zero at the origin and excludes a total of one
electron worth of charge in order to correct for self-interaction terms.
In contrast, the correlation hole integrates to zero and involves only a
redistribution of charge \cite{ch_bba89}.

A number of intuitive schemes for treating the exchange and correlation
holes have been proposed, where the pair correlation functions are
determined from a Schrodinger equation for an electron pair with
appropriate boundary conditions.\cite{kimball,bjhole,over95,jarl99,gori01}
Such models are useful for generalizing the LDA
by improving the description of exchange and correlation
in inhomogeneous systems.
Note, however, that in going beyond the LDA, a distinction should
be made between theories which attempt to find better energy functionals
but still lie within the framework of the DFT, and those theories which
focus on the self-energy $\Sigma_{xc}$ in order to obtain better quasiparticle
energies. The present scheme belongs to the
latter category where the excitation energies are given by
\begin{equation}
E_n=\varepsilon_n+\Delta_n~.
\label{eq_en}
\end{equation}
Here $\varepsilon_n$ is the eigenvalue corresponding to the Kohn-Sham
orbital $\psi_n$, and $\Delta_n$ is a state-dependent correction
associated with the self-energy $\Sigma_{xc}$. By using a particular
ansatz for $\Sigma_{xc}$, we have previously proposed\cite{bba04}
\begin{equation}
\Delta_n=
\int d^3 {\bf r}~
(2\epsilon_{xc}({\bf r})-v_{xc}({\bf r}))~|\psi_n({\bf r})|^2~,
\label{eq_sigman}
\end{equation}
where
\begin{equation}
v_{xc}({\bf r})=\frac{\delta E_{xc}}{\delta n({\bf r})}~,
\end{equation}
is the Kohn-Sham potential and
\begin{equation}
E_{xc}=\int d{\bf r}~n({\bf r})\epsilon_{xc}({\bf r})
\end{equation}
is the exchange-correlation energy. The exchange-correlation energy per
particle $\epsilon_{xc}$ can be expressed in terms of the pair correlation
functions as
\begin{equation}
\epsilon_{xc}({\bf r})=
\frac{1}{2}
\int d{\bf r'} \int_0^1 d\lambda
\frac
{n({\bf r'})C_{\lambda}({\bf r},{\bf r'})}
{|{\bf r}-{\bf r'}|},
\label{exc}
\end{equation}
where $\lambda$ is the coupling constant from the Hellmann-Feynman
theorem. Within the LDA, $\epsilon_{xc}({\bf r})$ is a well-known function
of the local density $n({\bf r})$ obtained from homogeneous electron gas
results \cite{review_dft}.

\section{spin resolved quasi-particle correction}

In the spin resolved case, the prescription for the correction to the
excitation energy given by Eq.~\ref{eq_sigman} becomes
\begin{equation}
\Delta_n^{\sigma}=
\int d^3 {\bf r}~
(2\epsilon_{xc}^{\sigma}({\bf r})-v_{xc}^{\sigma}({\bf r}))~
|\psi_n^{\sigma}({\bf r})|^2~,
\label{eq_sigmans}
\end{equation}
where
\begin{equation}
v_{xc}^{\sigma}({\bf r})=
\frac{\delta E_{xc}}{\delta n^{\sigma}({\bf r})}~,
\end{equation}
is the spin-dependent DFT potential and $n_{\sigma}$ the spin-resolved
electron density. As in Eq.~11 above, the exchange-correlation energy per
particle $\epsilon_{xc}^{\sigma}({\bf r})$ is given by the Coulomb
interaction of an electron of spin $\sigma$ at ${\bf r}$ with its
exchange-correlation hole:
\begin{equation}
\epsilon_{xc}^{\sigma}({\bf r})=
\frac{1}{2}
\sum_{\sigma'}
\int d{\bf r'} \int_0^1 d\lambda
\frac{
n_{\sigma'}({\bf r'})
C_{\lambda}^{\sigma,\sigma'}({\bf r},{\bf r'})}
{|{\bf r}-{\bf r'}|}~,
\label{excs}
\end{equation}
in terms of the spin-resolved correlation functions
$C_{\lambda}^{\sigma,\sigma'}({\bf r},{\bf r'})$. It is convenient to
split $\epsilon_{xc}^{\sigma}({\bf r})$ as
\begin{equation}
\epsilon_{xc}^{\sigma}({\bf r})=
\epsilon_{x}^{\sigma}({\bf r})+\epsilon_{c}^{\sigma}({\bf r}),
\end{equation}
where $\epsilon^{\sigma}_{x}$ and $\epsilon^{\sigma}_{c}$ give the
exchange and correlation components of $\epsilon_{xc}^{\sigma}({\bf r})$,
respectively. The spin-resolution of $\epsilon_x^{\sigma}$ is
straightforward because the exchange hole only involves like spin
electrons. On the other hand, the spin-resolution of $\epsilon_c^{\sigma}$
is more subtle since here both like and unlike spin electrons contribute.
The energy per particle for the spin polarized HEG depends only
on the density parameter $r_s$ and the local spin polarization
\begin{equation}
\zeta=(n_{\uparrow}-n_{\downarrow})/n~.
\end{equation}

In this connection, the spin-resolution of the correlation energy
presented recently by Gori-Giorgi and Perdew \cite{gori04} for the HEG may 
usefully be
employed. Ref.~\onlinecite{gori04} defines spin resolved correlation
energies
\begin{equation}
\epsilon_{c}^{\sigma \sigma'}(r_s,\zeta)=
2\pi\frac{n_{\sigma}}{n} \int_0^{\infty}
n_{\sigma'}{\bar C}_{c}^{\sigma,\sigma'}
(r_s,\zeta,u)~udu~,
\label{ecssp}
\end{equation}
by using correlation holes $n_{\sigma'}{\bar C}_{c}^{\sigma,\sigma'}
(r_s,\zeta,|{\bf r}-{\bf r'}|)$, where the pair correlation function
${\bar C}_{c}^{\sigma,\sigma}$ has been averaged over the coupling
strength $\lambda$. Here we have made the dependence of various quantities
on $r_s$ and $\zeta$ explicit. Thus the interaction energy
$\epsilon_{c}^{\sigma}$ of an electron of spin $\sigma$ with its
correlation hole can be resolved as
\begin{equation}
\epsilon_{c}^{\sigma}=\frac{n}{n_{\sigma}}\left (
\epsilon_{c}^{\sigma,\sigma}+
\frac{1}{2}\epsilon_{c}^{\sigma,-\sigma} \right ),
\label{ecs}
\end{equation}
Parameterized expressions for $\epsilon_{c}^{\sigma,\sigma}$ and
$\epsilon_{c}^{\sigma,-\sigma}$ based on the HEG are given in Ref.
\onlinecite{gori04}.

\begin{figure}
\begin{center}
\includegraphics[width=\hsize]{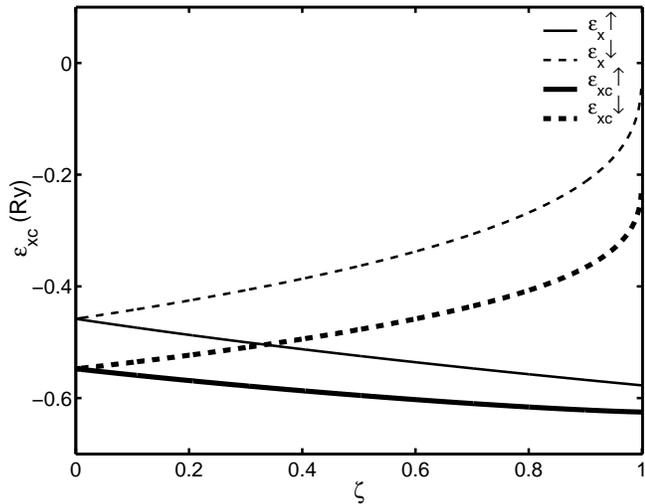}
\end{center}
\caption{
Characteristic behavior of the spin-resolved exchange-correlation energies 
$\epsilon_{xc}^{\uparrow}$ and $\epsilon_{xc}^{\downarrow}$ and the 
corresponding spin only contributions $\epsilon_{x}^{\uparrow}$ and 
$\epsilon_{x}^{\downarrow}$ as a function of the spin polarization $\zeta$ 
for $r_s=2$ in a HEG based on Eqs. 15-18. See 
legend for the meaning of various line types.}

\label{fig1}
\end{figure}

\begin{figure}
\begin{center}
\includegraphics[width=\hsize]{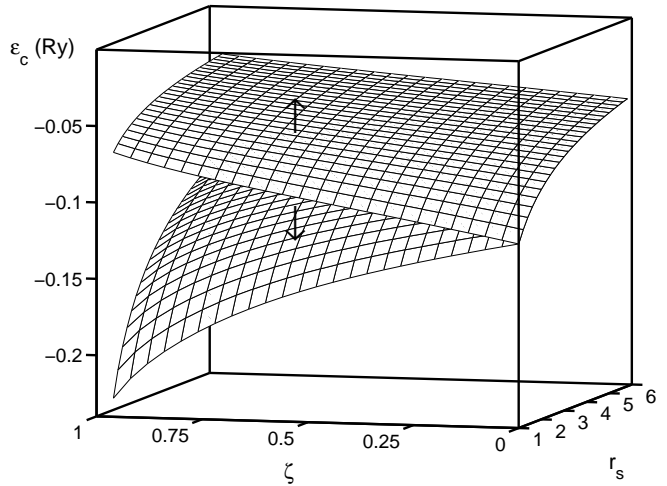}
\end{center}
\caption{
Spin-resolved correlation energies $\epsilon_{c}^{\sigma}(r_s,\zeta)$ (see 
Eq. 18) in a HEG are shown in surface plots as a 
function of the spin polarization $\zeta$ and the density parameter $r_s$. 
Upper and lower surfaces refer to the majority ($\uparrow$) and minority 
($\downarrow$) spins, respectively.
}
\label{fig2}
\end{figure}

Figs.~\ref{fig1} and ~\ref{fig2} present illustrative results concerning 
the $r_s$ and $\zeta$ dependencies of the exchange and correlation 
energies in the HEG. Fig.~\ref{fig1} shows the behavior of 
$\epsilon_{xc}^{\sigma}$ as a function of $\zeta$ for $r_s=2$. The 
correlation contribution $\epsilon_{c}^{\sigma}$ here has been obtained 
from Eq.~\ref{ecs}. The spin-splitting of $\epsilon_{xc}^{\sigma}$ is seen 
from Fig.~\ref{fig1} to increase steadily with $\zeta$. The exchange 
contribution at a given density depends linearly on 
$(1\pm \zeta)^{1/3}$ so that when $\zeta$ is small $\epsilon_x^{\sigma}$ displays a 
linear behavior with $\zeta$. Notably, $\epsilon_x^{\sigma}$ for minority 
spin electrons is seen in Fig.~\ref{fig1} (thin dashed line) to vanish in the 
ferromagnetic limit ($\zeta=1$) reflecting the fact that a minority spin 
electron has no other electrons to exchange with in this case. In 
contrast, the correlation energy per particle of the minority spin 
electrons does not vanish in the limit $\zeta=1$ due to the contribution 
of opposite spin electrons in Eq.~\ref{ecs}. The effect of correlations is to 
reduce the $\zeta$ dependency of $\epsilon_x^{\sigma}$. As a result, in the 
linear region, the slopes of $\epsilon_{xc}^{\sigma}$ curves are seen to be 
smaller and magnetic splittings are significantly reduced.

In order to further highlight the role of correlations, we show 
$\epsilon_{c}^{\uparrow}$ (upper surface) and $\epsilon_{c}^{\downarrow}$ 
(lower surface) as a function of the parameters $\zeta$ and $r_s$ in 
Fig.~\ref{fig2}. It is clear immediately that correlations yield a less 
attractive hole for the majority spins compared to that for the minority 
spins. The reason is that correlation effects are dominated by the 
contribution of opposite spin electrons which are not prevented by the 
Pauli principle to come close to each other. The splitting of the two 
surfaces increases with increasing spin polarization $\zeta$ but decreases 
with increasing $r_s$. At $\zeta=0$ the two surfaces collapse on to the 
well known paramagnetic curve, which scales as \cite{ch_bba89}
\begin{equation}
\epsilon_c=- \frac{0.127}{\sqrt{r_s}}~\mbox{Ry}
\end{equation}
for typical metallic densities ($r_s=1 - 6$).

\section{A More General Formulation}

This section presents a framework for generalizing Eq.~\ref{eq_en}. This
formulation also allows us to connect our approach to some related methods
in the literature. In the interest of notational simplicity, we will
omit from hereon the obvious spin dependencies and the associated
spin-indices.

We recall that the standard definition of the the self-energy operator
$\Sigma_{xc}$ is:
\begin{equation}
\Sigma_{xc}= G_0^{-1}-G^{-1}~,
\end{equation}
where $G$ is the exact single-particle Green's function and $G_0$ is the
reference Green's function obtained within the Hartree approximation. Thus
$\Sigma_{xc}$ in principle contains all quasiparticle effects beyond the
Hartree approximation. We now write $\Sigma_{xc}$ (suppressing
spin-indices) as
\begin{equation}
\Sigma_{xc}=
v_{xc}+
\sum_n \Delta_n |\psi_n><\psi_n|~,
\label{eq_sigma}
\end{equation}
where $v_{xc}$ is the spin-dependent exchange-correlation potential
associated with the spin-dependent Kohn-Sham orbitals $\psi_n$.
\cite{footnote_dyson}
The state-dependent correction $\Delta_n$ is more general
than the expression of Eq.~\ref{eq_sigmans} and it can be the basis for
curing the unphysical
interaction of an electron with itself which occurs in the DFT. In fact,
the right side of Eq.~\ref{eq_sigmans} attempts to correct for such
self-interactions by using the Coulomb potential of the
exchange-correlation hole, but does so only approximately since it assumes
every electron to possess the same pair-correlation function. In systems
with localized orbitals, it makes sense to subtract out the
self-interaction for each occupied orbital explicitly by using\cite{footnote_eq22}
\begin{equation}
\Delta_n=
\int d^3 {\bf r}~
(-u_n({\bf r})-v_{xc}^{n}({\bf r}))~
|\psi_n({\bf r})|^2~,
\label{eq_sic}
\end{equation}
where $u_n({\bf r})$ and $v_{xc}^{n}({\bf r})$ are the electrostatic and
the exchange-correlation potential, respectively, created by the charge
density $|\psi_n({\bf r})|^2$. 
The self-interaction correction (SIC) given
by Eq.~\ref{eq_sic} has been proposed by Perdew and Zunger \cite{sic} for
atomic systems and possesses a straightforward spin-dependence. For atoms,
SIC lowers the LDA eignevalues and improves agreement with
measurements\cite{review_GW}. Note, however, that in extended systems SIC
encounters conceptual problems due to the delocalized
character of the Bloch functions.\cite{norman84} In fact, the SIC 
correction turns to be
zero for plane-waves. A route around this problem is to renormalize the
wavefunction within the unit cell and perform the correction within the
atomic spheres.\cite{epmdlmto}

The so-called LDA+U can be considered a simplified
version of the SIC where the correction $\Delta_n^{\sigma}$ acts only on
localized $d$-or $f$-states, often splitting the metallic LDA bands into
upper and lower Hubbard bands descriptive of Mott-Hubbard insulators 
\cite{ldau,ldau_review,cococcioni}.
However, problems with both SIC and the LDA+U arise for systems with
partially filled 3$d$ shells which are metallic, as for example in the
transition metals.\cite{review_GW,cococcioni} In order to handle cases where a
conduction band crosses the Fermi energy $E_F$, we observe first that here
the DFT potential $v_{xc}$ generally provides a reasonable approximation
to the self-energy at the Fermi surface (FS) \cite{gunnar76} and we may
impose the condition
\begin{equation}
\Sigma_{xc}|_{FS}=
v_{xc}
\label{eq_sigmaf}
\end{equation}
on the self-energy. We now write $\Delta_n$ as
\begin{equation}
\Delta_n={\bar \Delta}+\Lambda_n~,
\label{eq_dec}
\end{equation}
where ${\bar \Delta}$ is defined to be the average correction on the FS
and ${\Lambda_n}$ is a state-dependent contribution. It is clear that the
condition of Eq.~\ref{eq_sigmaf} can be satisfied by setting the average
shift ${\bar \Delta}$ to zero and requiring that $\Delta_n$ goes to zero at
the $E_F$, which leads us to suggest the form
\begin{equation}
\Lambda'_n=\Lambda_n\frac{\varepsilon_n-E_F}{E_B-E_F}~,
\label{eq_met}
\end{equation}
in metallic systems. Here $E_B$ denotes the bottom of the conduction band.
The linear interpolation given by Eq.~\ref{eq_met} is equivalent to a band
renormalization near the Fermi level given by
\begin{equation}
{\tilde \xi_n}=(1+\lambda_n)\xi_n~,
\end{equation}
where
\begin{equation}
\lambda_n=\frac{\Lambda_n}{E_B-E_F}~
\label{eq_lam}
\end{equation}
and $\xi_n=(\varepsilon_n-E_F)$. Such a renormalization due to self-energy
effects is often invoked phenomenologically within the framework of Fermi
liquid theories.\cite{janak74}

\begin{figure}
\begin{center}
\includegraphics[width=\hsize]{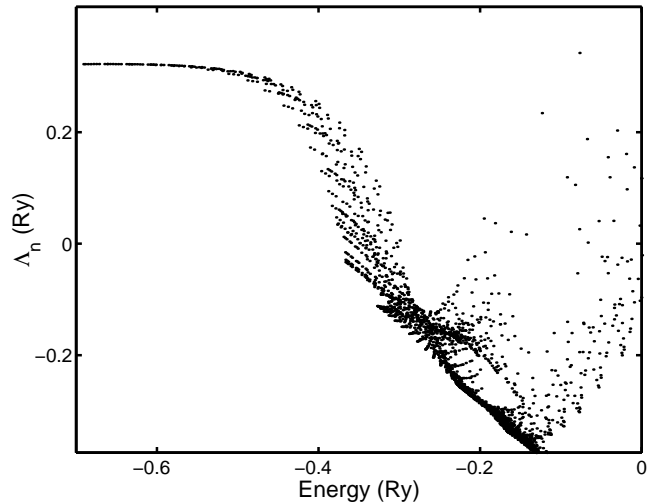}
\end{center}
\caption{
State-dependent correction $\Lambda_n$ based on Eqs.~\ref{eq_dec} and \ref{eq_sic} 
in Cu as a function of energy with zero denoting the Fermi energy. 
Each dot gives the value of $\Lambda_n$ for a particular state. 
Total number of states shown is $6\times 505$ corresponding to $6$ valence bands 
at each of the $505$ k-points computed in the 1/48th irreducible portion of 
the Brillouin zone of Cu as discussed in the text. 
}
\label{fig3}
\end{figure}

Fig.~\ref{fig3} shows illustrative results in Cu where the electronic
structure involves states of a mixed $s-p$ and $d$ character. 
For this purpose, the standard LDA-based energy bands and Bloch 
wavefunctions were computed for
the $6$ valence bands of fcc Cu on a uniform $505$ $\bf k$-point mesh in the
irreducible 1/48th of the Brillouin zone. The correction $\Delta_n$ of
Eqs.~\ref{eq_sic} and \ref{eq_dec} 
was then computed for each of the 505$\times$6 states
\cite{footnote_deltan}. 
The resulting values of $\Lambda_n$ after the average $\bar\Delta$ is
subtracted are shown in the 'dot-plot' of Fig.~\ref{fig3}. 
$\Lambda_n$ is seen to be positive and fairly constant at 
low band energies ($-0.7$ to $-0.5$
Ry), where the states mainly possess $s$ character. 
At intermediate energies ($-0.4$ to $-0.2$ Ry),
where the $d$ admixture increases, $\Lambda_n$ decreases rapidly as
self-interaction effects become stronger. Closer to the Fermi energy
($-0.1$ to $0.0$ Ry), the $s-p$ character increases once again and this is
reflected in the upturn in $\Lambda_n$ values seen in Fig.~\ref{fig3}. 
Using $E_F-E_B=0.691$ Ry, we find that near the FS, typical 
$\lambda_n$ values are $\approx 10$ \%, which is consistent with 
the value of the renormalization parameter $\lambda\approx 8$ \% 
adduced from fitting experimental
photoemission results \cite{norman84,janak74,footnote_norman}.

In ferromagnetic metals such as Fe and Ni the present scheme leaves the 
band structure at the Fermi energy and hence the majority and minority 
spin FS's and the corresponding spin magnetic moments unchanged due to the 
constraint of Eq.~\ref{eq_sigmaf}. However, energy bands away from the 
Fermi level are modified. In particular, the topmost occupied $d$ states 
in 3$d$ and 4$d$ transition metals, which are more localized within the 
atomic sphere than the $d$ states toward the bottom of the band, will 
experience larger corrections (lowering). As a result, $d$-band widths 
would become narrower giving trends consistent with photoemission 
experiments \cite{eastman78}.

\section{summary and conclusions}

We discuss how DFT eignevalues can be corrected via a spin-dependent 
self-energy $\Sigma_{xc}^{\sigma}$ to obtain improved quasi-particle energies. 
For this 
purpose, $\Sigma_{xc}^{\sigma}$ is evaluated as the Coulomb 
energy associated with the 
exchange and correlation holes surrounding each electron in terms of the 
appropriate spin-resolved pair correlation functions for the homogeneous 
electron gas. The approach is similar to that followed in our earlier 
treatment of the unpolarized case \cite{bba04}, and should yield 
quasi-particle energies with an accuracy comparable to that of the more 
demanding GWA. Notably, our scheme does not involve the 
limitations of the standard implementations of the GWA in 
which the screened interaction $W$ is assumed to be spin-independent. We 
also show how our scheme can be viewed in a somewhat more general context, 
so that quasi-particle energies can be corrected on a state-by-state basis 
for self-interaction terms in the DFT along the lines of the familiar SIC 
scheme \cite{sic}. The question of applying self-interaction 
correction to a metallic band is considered and an interpolation formula 
to handle such a case is suggested. Finally, illustrative results in Cu 
are presented and the computed renormalization of the DFT eignvalues of 
the conduction band in Cu is found to be in reasonable accord with the 
corresponding experimental results.

\acknowledgements

We acknowledge Paola Gori-Giorgi, Hsin Lin and Robert Markiewicz for 
useful discussions. This work is supported by the US Department of Energy 
contract DE-AC03-76SF00098 and benefited from the allocation of 
supercomputer time at NERSC and Northeastern University's Advanced 
Scientific Computation Center (ASCC).


\begin{thebibliography}{99}
%
%
\bibitem{review_GW}
F. Aryasetiawan and O. Gunnarsson,
Rep. Prog. Phys. {\bf 61}, 237 (1998);
see also Y. Kubo, J. Phys. and Chem. of Solids,
present volume.
%
%
\bibitem{bba04}
B. Barbiellini and A. Bansil,
J. Phys. and Chem. of Solids {\bf 65}, 2031(2004).
%
%
\bibitem{review_dft}
R.O. Jones and O. Gunnarsson,
Rev. Mod. Phys {\bf 61}, 689 (1989).
%
%
\bibitem{fritsche86}
L. Fritsche, Phys. Rev. B {\bf 33}, 3976 (1986).
\bibitem{fritsche93}
L. Fritsche and Y. M. Guo,
Phys. Rev. B {\bf 48}, 4197 (1993).
\bibitem{freeman94}
D. Vogtenhuber, R. Podloucky, A. Neckel, S. G. Steinemann,
and A. J. Freeman,
Phys. Rev. B {\bf 49}, 2099 (1994).
\bibitem{remed99}
I. N. Remediakis and E. Kaxiras,
Phys. Rev. B {\bf 59}, 5536 (1999).
\bibitem{reed00}
E. J. Reed, J. D. Joannopoulos, and L. E. Fried,
Phys. Rev. B {\bf 62}, 16500 (2000).
\bibitem{sankey00}
J. Dong, O. F. Sankey, S. K. Deb, G. Wolf,
and P. F. McMillan,
Phys. Rev. B {\bf 61}, 11979 (2000).
\bibitem{sankey01}
M. Fuentes-Cabrera and O. F. Sankey,
J.Phys.: Condens. Matter {\bf 13}, 1669 (2001).
%
\bibitem{hfs}
J. C. Slater, Phys. Rev. {\bf 81}, 385 (1951).
\bibitem{ks}
W. Kohn and L. J. Sham, Phys. Rev. {\bf 140}, A1133 (1965).
%
%
\bibitem{ch_bba89}
B. Barbiellini, Phys. Lett. A {\bf 134}, 330 (1989).
%
%
\bibitem{kimball}
J.C. Kimball, Phys. Rev. A {\bf 7}, 1648 (1973).
\bibitem{bjhole}
B. Barbiellini and T. Jarlborg,
J.Phys.: Condens. Matter {\bf 1}, 75 (1989).
\bibitem{over95}
A.W. Overhauser, Can. J. Phys. {\bf 73}, 683 (1995).
\bibitem{jarl99}
T. Jarlborg, Phys. Lett. A {\bf 260}, 395 (1999).
\bibitem{gori01}
P. Gori-Giorgi and J.P. Perdew, Phys. Rev. B
{\bf 64}, 155102 (2001);
M. Corona, P. Gori-Giorgi, and J. P. Perdew,
Phys. Rev. B {\bf 69}, 045108 (2004).
%
%
\bibitem{gori04}
P. Gori-Giorgi and J. P. Perdew,
Phys. Rev. B {\bf 69}, 041103(R) (2004);
Spin-resolved code for obtaining correlation 
energy was provided
by P. Gori-Giorgi.
%
%
\bibitem{footnote_dyson}
The second term in Eq.~\ref{eq_sigma} should more properly contain an
expression in terms of the Dyson orbitals
(see, e.g. Ref. \onlinecite{bba04}).
However, the Kohn-Sham orbitals are generally expected to provide a
reasonable approximation to the Dyson orbitals
(see, e.g. P. Duffy, D. P. Chong, M. E. Casida, and D. R. Salahub,
Phys. Rev. A {\bf 50}, 4707 (1994)).
%
%
\bibitem{footnote_eq22}
The exchange hole defined by Slater (Ref.~\onlinecite{hfs}) depends
on the quantum state $\psi_n$ of the electron located at ${\bf r}_1$.
In a system with extended states, this hole is not very different from
the hole averaged over all occupied states,
but in a system with localized orbitals,
the exchange-correlation hole strongly
depends on $\psi_n$. In the latter case, the main
effect of the exchange-correlation hole 
is to cancel the self-interaction terms 
produced by the electron
in the state $\psi_n$.
\bibitem{sic}
J. P. Perdew and A. Zunger,
Phys. Rev. B {\bf 23}, 5048 (1981).
\bibitem{norman84}
M.R. Norman,
Phys. Rev. B {\bf 29}, 2956 (1984).
%
%
\bibitem{epmdlmto}
B. Barbiellini, S. B. Dugdale and T. Jarlborg,
Computational Materials Science {\bf 28}, 287 (2003).
%
%
\bibitem{ldau}
V. I. Anisimov, J. Zaanen, and O. K. Andersen,
Phys. Rev. B {\bf 44}, 943 (1991).
\bibitem{ldau_review}
V. I. Anisimov, F. Aryasetiawan and A.I. Lichtenstein,
J. Phys.: Condens. Matter {\bf 9}, 767 (1997).
\bibitem{cococcioni}
M. Cococcioni and S. de Gironcoli,
Phys. Rev. B {\bf 71}, 035105 (2005).
%
%
\bibitem{gunnar76}
O. Gunnarsson and B.I. Lundqvist,
Phys. Rev. B {\bf 13}, 4274 (1976).
%
%
\bibitem{janak74}
J.F. Janak, A.R. William and V.L. Moruzzi,
Phys. Rev. B {\bf 11}, 1522 (1975).
%
\bibitem{footnote_deltan}
Eqs.~\ref{eq_sigman} and \ref{eq_dec} 
give $\Lambda_n$ values for Cu 
which are about 30\% larger for 
the topmost occupied 
$d$-like states.
\bibitem{footnote_norman}
A different SIC calculation in Cu
(Ref. \onlinecite{norman84}) also
suggests that SIC should yield eigenvalues close to experimental
excitation energies in Cu and other transition metals when 
appropriate modifications for metals are implemented.
%
%
\bibitem{eastman78}
D. E. Eastman, F. J. Himpsel, and J. A. Knapp, 
Phys. Rev. Lett. {\bf 40}, 1514 (1978).
\end{thebibliography}
\end{document}